\documentclass[aip,rsi,amsmath,amssymb,reprint,floatfix,10pt]{revtex4-2}

\usepackage{mathptmx}

\usepackage{times}

\usepackage{amssymb,amsmath}
\usepackage{graphicx}
\usepackage{natbib}
\usepackage{multirow}
\usepackage{float}
\usepackage{braket}
\usepackage{silence}
\usepackage[dvipsnames]{xcolor}
\usepackage[version=4]{mhchem}
\usepackage{soul}
\begin{document}

\title{Surface Chemistry-Driven Oxidation Mechanisms in Ti$_{\text{3}}$C$_{\text{2}}$T$_{\textit{x}}$ MXenes}

\author{Bradlee~J.~McIntosh}
\affiliation{Stavropoulos Center for Complex Quantum Matter, Department of Physics and Astronomy, University of Notre Dame, Notre Dame, Indiana 46556, USA}

\author{Bence~G.~M\'arkus}
\affiliation{Stavropoulos Center for Complex Quantum Matter, Department of Physics and Astronomy, University of Notre Dame, Notre Dame, Indiana 46556, USA}

\author{Anna~Ny\'ary}
\affiliation{Stavropoulos Center for Complex Quantum Matter, Department of Physics and Astronomy, University of Notre Dame, Notre Dame, Indiana 46556, USA}

\author{Ferenc~Simon}
\affiliation{Stavropoulos Center for Complex Quantum Matter, Department of Physics and Astronomy, University of Notre Dame, Notre Dame, Indiana 46556, USA}
\affiliation{Department of Physics, Institute of Physics and ELKH-BME Condensed Matter Research Group Budapest University of Technology and Economics, M\H{u}egyetem rkp. 3., H-1111 Budapest, Hungary}
\affiliation{Institute for Solid State Physics and Optics, Wigner Research Centre for Physics, Budapest H-1525, Hungary}

\author{L\'aszl\'o~Forr\'o}
\email[Corresponding author: ]{lforro@nd.edu}
\affiliation{Stavropoulos Center for Complex Quantum Matter, Department of Physics and Astronomy, University of Notre Dame, Notre Dame, Indiana 46556, USA}

\author{D\'avid~Beke}
\affiliation{Stavropoulos Center for Complex Quantum Matter, Department of Physics and Astronomy, University of Notre Dame, Notre Dame, Indiana 46556, USA}
\affiliation{Institute for Solid State Physics and Optics, Wigner Research Centre for Physics, Budapest H-1525, Hungary}

\begin{abstract}
    Ti$_3$C$_2$T$_x$ is a leading compound within the MXenes family and can find host in widespread applications. It is synthesized by selectively etching layers from the Ti$_3$AlC$_2$ precursor, and this process typically introduces surface terminations, T$_x$, such as \ce{\bond{-}OH}, \ce{\bond{=}O}, or \ce{\bond{-}F}. However, the aggressive chemical conditions required for its preparation, as well as exposure to air, humidity, and heat, can lead to impurity phases that potentially compromise its desirable properties. We reveal a two-step oxidation process during heat treatment, where initial oxidation occurs between layers without altering the integrity of the Ti$_3$C$_2$ layered structure, followed by the formation of anatase TiO$_2$ at elevated temperatures. The process was carefully monitored using \emph{in situ} Raman spectroscopy and \emph{in situ} microwave conductivity measurements, employed to Ti$_3$C$_2$T$_x$ prepared using various etching techniques involving concentrated HF, {\ce{LiF + HCl}}, and \ce{HF + HCl} mixtures. The oxidation process is heavily influenced by the synthesis route and surface chemistry of Ti$_3$C$_2$T$_x$, with fluoride and oxyfluoride groups playing a pivotal role in stabilizing the anatase phase. The absence of these groups, in contrast, can lead to the formation of rutile
    TiO$_2$. %These findings elucidate a method for controlled surface oxidation and modification to optimize Ti$_3$C$_2$T$_x$ properties for tailored applications.
\end{abstract}

\maketitle

\section{Introduction}

Research into novel 2D materials, such as MXenes is beneficial to enable and improve novel devices that are otherwise inaccessible using conventional materials, e.g., wearable, flexible electronics \cite{LyuACSNano2019, MaAFM2021}, thanks to their unique atomic-layered structures and tunable properties. These materials hold great promise for advancing technologies in energy storage, electronics, catalysis, and environmental remediation.

MXenes are formed by selectively etching the A layers from the MAX phases \cite{barsoum_max_2013}, leading to a structure of M$_{n+1}$X$_n$T$_x$, where M is an early transition metal (e.g., titanium, vanadium, chromium, molybdenum), A is an element, primarily from groups 13 and 14 (e.g., aluminum, silicon, or gallium), and X is carbon, nitrogen, or both. Furthermore, $n$ can be $1$, $2$, or $3$ and T$_x$ represents surface terminations that are typically \ce{\bond{-}OH}, \ce{\bond{=}O}, or \ce{\bond{-}F}. These materials have emerged as leading candidates for a multitude of advanced technological applications due to their tunable electrical, mechanical and surface properties.\cite{riss_stability_2009, chaturvedi_rise_2023, tahir_introduction_2024} The potential to tailor these, positions MXenes at the focus of research across diverse fields, such as energy storage, catalysis, and electromagnetic interference shielding. \cite{fatima_ferroelectric-controlled_2023, jiang_2d_2022, tahir_introduction_2024} and they may also find applications in the field of nanoantenna based sensorics \cite{szakmany_angle_2024, szakmany_long-wave_2024}  because of their exceptionally good microwave and THz absorption capabilities. However, one of the most significant challenges limiting their widespread adoption is their inherent susceptibility to oxidation, which can severely compromise their structural integrity and reduce their performance even under ambient conditions and especially under elevated humidity.\cite{xia_ambient_2019, habib_oxidation_2019, zhang_oxidation_2017, soomro_progression_2023}

The oxidation of MXenes, particularly for Ti$_3$C$_2$T$_x$, typically results in the formation of titanium dioxide (TiO$_2$), which detrimentally alters their properties by increasing electrical resistivity and reducing the number of functional surface sites.\cite{zhang_oxidation_2017, cao_recent_2022} Understanding the mechanisms driving this oxidative degradation is therefore critical for developing strategies to enhance MXene stability. Previous studies have demonstrated that oxidation is influenced by various environmental factors such as exposure to air,\cite{xu_high-temperature_2006} moisture,\cite{liu_mechanism_2024} and light,\cite{zhang_oxidation_2017} with oxidation often initiating at defect sites or edges where TiO$_2$ nanoparticles preferentially form. A recent theoretical study demonstrated that the nature of the surface terminations—such as \ce{-H}, \ce{-OH}, and \ce{-O} significantly influences the interaction strength between Ti$_2$CT$_x$ surface and the formed TiO$_2$.\cite{garcia_effect_2024}

Nonetheless, the controlled oxidation of MXenes has been shown to confer beneficial properties in certain contexts. For instance, the formation of specific TiO$_2$ polymorphs, such as anatase and rutile, can enhance the electrochemical performance of MXenes in supercapacitors and batteries.\cite{cao_room_2017, dong_ti3c2_2017} In both applications, phase purity of the formed oxide is a crucial factor. Once a connected crystalline film of TiO$_2$ is formed on the surface of the Ti$_3$C$_2$, one can differentiate two possible scenarios. i) Anatase TiO$_2$ has a crystal structure consisting of distorted edge-sharing TiO$_6$ octahedra. This structure features a three-dimensional zig-zag pathway in the crystal lattice, allowing Na$^+$ and Li$^+$ to diffuse along various directions.\cite{li_sodium-ion_2022, steele_anisotropy_1969}. ii) On the other hand, rutile TiO$_2$ forms one-dimensional ion diffusion paths along its $c$-axis. As an anode material, rutile is considered to have lower electrochemical activity, however, the Li$^+$ diffusion coefficient along the $c$-axis in rutile TiO$_2$ is extremely high (in the order of $10^{-6}$~cm$^2$/s)\cite{steele_anisotropy_1969} that gives direction selectivity for ion transport.

Moreover, the possible presence of titanium suboxides and Magn\'eli phases offers intriguing possibilities for applications requiring mixed valence states and high electrical conductivity on the surface,\cite{tahir_first_2022, sattar_ferroelectric_2024} which is also promoted by the already high conductivity of the underlying MXene structure. Given the critical importance of the TiO$_2$ phase in these applications, a detailed understanding of the oxidation process, as well as the ability to control it, is essential for optimizing the properties of MXene-based materials and compositions.

In order to gain further insights into the oxidation processes, we investigated Ti$_3$C$_2$T$_x$ synthesized through various etching routes, focusing on the influence of heating by either using a laser beam or introducing direct heating. The evolution of the material was monitored by \emph{in situ} Raman spectroscopy using laser-induced heating and verified by following the stepwise oxidation in \emph{in situ} microwave conductivity measurements applying a direct heat flow. Recently, the first approach was used to monitor the oxidation of Ti$_3$C$_2$T$_x$ MXenes synthesized via the LiF-HCl etching.\cite{adomaviciute-grabusove_monitoring_2024} We, however, focus on Ti$_3$C$_2$ prepared by different routes and monitored the beginning of the oxidation carefully to reveal its reaction kinetics. Our findings reveal a two-step oxidation mechanism and TiO$_2$ phase affinity that is highly dependent on the conditions of preparation and surface chemistry, with significant implications for the controlled synthesis of TiO$_2$@C composites and the stabilization of specific TiO$_2$ polymorphs.

\section{Results and Discussion}
\paragraph{Synthesis and characterization}

\begin{figure}[htp]
    \centering
    \includegraphics[width=1\linewidth]{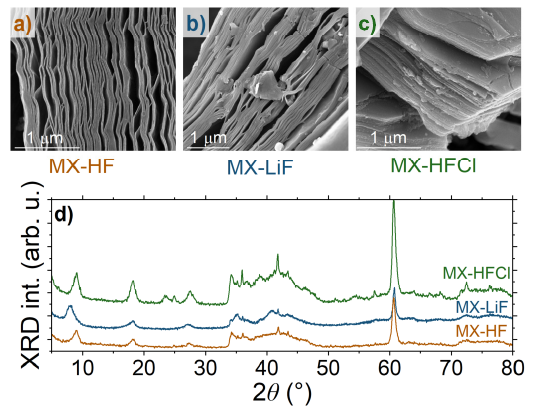}
    \caption{(a-c) SEM images and (d) XRD data of Ti$_3$C$_2$T$_x$ synthesized through three different methods: concentrated HF ({\color{brown}MX-HF}), LiF-HCl solution ({\color{MidnightBlue}MX-LiF}), and HF-HCl mixture ({\color{OliveGreen}MX-HFCl}). The scale bars in the SEM images are all $1$ $\mu$m.} 
    \label{XRD}
\end{figure}

For the synthesis of Ti$_3$C$_2$T$_x$ MXenes, three different approaches, adapted from previous reports\cite{benchakar_one_2020, alhabeb_guidelines_2017}, were used: etching in concentrated HF ($48\%$, "MX-HF"), LiF-HCl solution ("MX-LiF"), and HF-HCl mixture ("MX-HFCl"). Delamination was also performed on a certain amount of MX-LiF material using ultrasound sonication ("MX-LiF del"). Characterizations of the resulting MXenes were carried out using X-ray diffraction (XRD), scanning electron microscopy (SEM), and elemental analysis, which confirmed successful etching and the formation of the characteristic accordion-like layered structure (Fig. \ref{XRD}). 

\paragraph{Effects of Annealing}
Signs of oxidation is apparent when the samples were annealed. Here, we present measurements on samples annealed at $400~^{\circ}$C in air or in vacuum for two hours to compare laser heating with conventional annealing. The samples annealed in air are labeled as "400-Air", and the vacuum-annealed samples are labeled as "400-Vac". The microscopic examination, Raman, and X-ray photoelectron spectroscopy (XPS) measurement showed differences between the samples after annealing based on the type of synthesis method used.

\begin{figure}[hbp]
    \centering
    \includegraphics[width=1\linewidth]{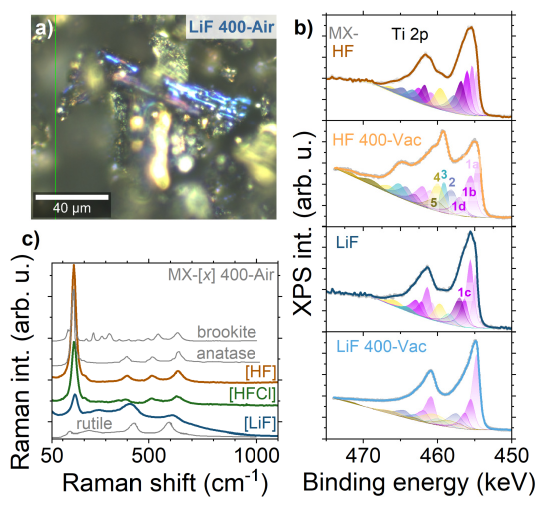}
    \caption{(a) Microscope image of an MX-LiF particle with layered oxidation after annealing in Air. (b) Ti 2p high-resolution XPS spectra of different MXenes before and after vacuum annealing. Notations follow the suggestion of Natu \emph{et al}.,\cite{natu_critical_2021}: \textbf{1a-d} correspond to the surface Ti atoms with $+1$ oxidation states and with three \ce{C\bond{-}Ti} bonds. The remaining possible bindings are O/O/O, O/O/F, O/F/F, and F/F/F, meaning that the Ti atom can bind to $1-3$ oxygen or fluorine atoms. Peaks labeled with \textbf{2}, \textbf{3} are Ti$^{2+}$ and Ti$^{3+}$, respectively, peaks \textbf{4} and \textbf{5} are Ti$^{4+}$. Peak \textbf{4} is dedicated to TiO$_2$, while \textbf{5} is Ti$^{4+}$, having \ce{Ti\bond{-}F} bonds. (c) Raman spectra of air-oxidized samples of all synthesis methods compared to the spectra of the rutile, anatase, and brookite phases of TiO$_2$.}
    \label{Fig. 2}
\end{figure}

Two primary conclusions can be drawn from these measurements. First, the MX-LiF sample demonstrates enhanced oxidation resistance compared to MX-HF and MX-HFCl. This is evidenced by the inhomogeneous oxidation patterns and layered structures observed in microscopic images (Fig.~\ref{Fig. 2}a)), as well as by the lower intensity of TiO$_2$-related features in the XPS spectra of vacuum-annealed specimens (Fig.~\ref{Fig. 2}b)). Indeed, deconvolution of the Ti 2p XPS peaks, based on the methodology proposed by Natu \emph{et al}.,\cite{natu_critical_2021} reveals the presence of multiple oxidation states, including Ti$^{3+}$ and Ti$^{4+}$. Notably, the vacuum-annealed samples also display distinct \ce{Ti\bond{-}F} bonds, and a significant reduction in highly fluorinated \ce{Ti\bond{-}C} bonds. Together, with increased peak intensities related to TiO$_2$ and \ce{TiO$_x$\bond{-}F} / TiF$_4$ for MX-HF 400-Vac, this suggests that annealing plays a crucial role in modifying surface chemistry and, consequently, the oxidation behavior.

Second, the TiO$_2$ phase that forms under annealing in air depends strongly on the etching route. HF etching predominantly yields anatase, whereas LiF-based etching favors rutile TiO$_2$. This conclusion is substantiated by the Raman spectra shown in Fig.~\ref{Fig. 2}c). For clarity, they are shown together with the three most common phases of TiO$_2$: rutile, anatase, and brookite. All samples, except MX-LiF 400-Air, contained exclusively the anatase phase of TiO$_2$. MX-LiF 400-Air, on the other hand, oxidized to a mixture of rutile and anatase TiO$_2$ phases, with greater structural integrity and lower background photoluminescence compared to other samples. 
%This is also hinted by the higher conductivity of the LiF-HCl etched material discussed in numerous articles.\cite{jia_tuning_2023, barsoum_electrical_2000} The reason behind the enhanced durability of the MX-LiF is probably due to the different ratio of surface functional groups, e.g., less \ce{\bond{-}F} bonds and/or more \ce{\bond{-}OH} is present than in the other samples.

\begin{figure}[h!tp]
    \centering
    \includegraphics[width=1\linewidth]{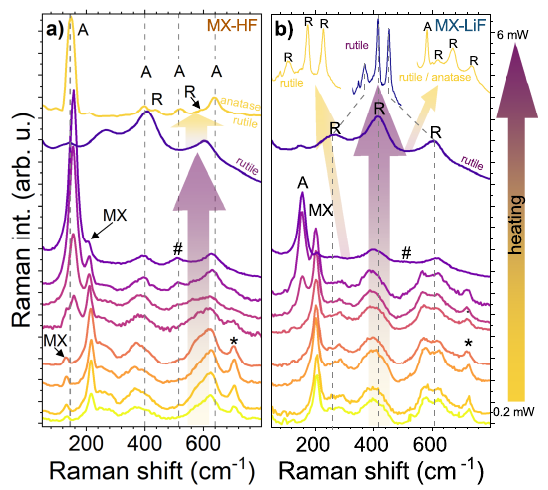}
    \caption{The laser-induced oxidation of MX-HF (a) and MX-LiF (b). MX-HF oxidizes to an anatase phase which transforms into rutlie at higher temperatures. The rutile phase transforms back to anatase when the sample cools down. MX-LiF on the other hand, transforms into rutile when it cools down or is heated further. Arrows with transition color indicate the temperature changes between measurements, heating or cooling. \# marks the position of the anatase $A_{1\text{g}}$ mode around $512$ cm$^{-1}$ and $\ast$ marks the disappearing MXene peak around $720$ cm$^{-1}$. Peak positions of Anatase and rutile TiO$_2$ are labeled as A and R, respectively}
    \label{figOX}
\end{figure}

\paragraph{Following the Oxide Growth In Situ}
Given that \emph{ex situ} characterization indicated that oxidation behavior is strongly influenced by the synthesis history, we conducted \emph{in situ} Raman and microwave conductivity measurements to gain deeper insight into the underlying oxidation mechanisms and to clarify the specific pathways governing the formation of various oxide phases. Figure \ref{figOX} presents the evolution of the Raman spectra of MX-HF and MX-LiF as the laser power was increased from $200~\mu$W ($18$~kW/cm$^2$) to $6$~mW ($550$~kW/cm$^2$), effectively heating the samples. Here, the optical irradiance, given in braces, is calculated based on the Airy disc diameter of $1.09~\mu\text{m}^2$. Oxidation typically begins at approximately $600~\mu$W ($55$~kW/cm$^2$) for MX-HF and MX-HFCl, and around $1$~mW ($92$~kW/cm$^2$) for MX-LiF, MX-LiF del, and MX-HF 400-Vac. Moreover, we can identify two distinguishable behaviors that exhibit similar features over all samples: (i) MX-HF, MX-HFCl, and (ii) MX-LiF, MX-LiF del, MX-HF 400-Vac, MX-LiF 400-Vac. Consequently, we focus our subsequent discussion on MX-HF, MX-HF 400-Vac, and MX-LiF. %, as $d = 1.22 \times \lambda/\mathrm{NA} = 1.18~\mu$m, where $\lambda=532$ nm is the excitation wavelength and $\mathrm{NA}=0.55$ is the numerical aperture of the used objective. The resulting spot size is thus $1.09~\mu\text{m}^2$ in all our measurements.

In the case of the MX-HF sample (group i), shown in Fig. \ref{figOX}a), the first observable change is the emergence of a new peak at 157~cm$^{-1}$, accompanied by a simultaneous decrease in the $720$~cm$^{-1}$ peak. As the laser power, and by extension the temperature as well, increased, the low-wavenumber peak red-shifted to $155$~cm$^{-1}$, and a new peak appeared at $512$~cm$^{-1}$. Although the changes in sample composition are evident in the Raman spectra from the emergence of the $512$~cm$^{-1}$ peak that closely resembles the spectral signature of anatase, unambiguously identifying a single oxide phase (or other underlying chemical, or physical transformations) remains challenging. Previous studies labeled all peaks around $155$ cm$^{-1}$ as the $E_{\text{g}}$ phonon mode of anatase TiO$_2$ in the context of MXene oxidation. However, this position represents a significant blue shift for this Raman mode, given that the first $E_{\text{g}}$ peak of crystalline anatase TiO$_2$ is reported in the literature at $144–147$ cm$^{-1}$.\cite{fujiki_jrs_1978,eror_jssc_1982} Various factors, such as tensile strain, doping, oxygen deficiency, and the presence of Ti$^{3+}$, can induce such a shift towards higher wavenumbers; nevertheless, most studies report blue-shifted $E_{\text{g}}$ values at lower wavenumbers~\cite{eror_jssc_1982, chen_jphysd_2000, balaji_phonon_2006, xu_blue_2001, iliev_raman_2013, liu_fabrication_2018, chen_black_2016}.

The other Raman-active modes of rutile TiO$_2$ are the $B_{1\text{g}}$ and $E_{\text{g}}$ modes at $400$ cm$^{-1}$ and $640$ cm$^{-1}$, of which both overlap with the Raman signals of Ti$_3$C$_2$T$_x$, and the $A_{1\text{g}}$ mode at around $515$ cm$^{-1}$. Although observing the latter (together with the intense $E_{\text{g}}$  mode) suggests the formation of rutile, similarities between the Raman spectra of suboxides, and the blue-shifted $E_{\text{g}}$ mode make it difficult to rule out the presence of a highly disturbed anatase phase, or the presence of suboxides, such as Ti$_2$O$_3$ or Ti$_3$O$_5$, or one of the Magn\'eli phases.\cite{malik_modelling_2020, el_marssi_ferroelectric_2003, wu_high_2012} at the beginning of the oxidation. As the temperature increased further, the oxide transforms into rutile. Once the laser-induced heating stopped, the rutile phase transformed to an oxide with a $E_{\text{g}}$ peak at $144$ cm$^{-1}$, which is typically attributed to anatase.

The MX-LiF (see Fig. \ref{figOX}b)) and vacuum-annealed samples behave quite differently. First, the low-wavenumber peak appeared around $154$ cm$^{-1}$ instead of $157$ cm$^{-1}$, without the previously identified peak shift changes at the beginning of the oxidation. In stark contrast to the evolution of the MX-HF sample, the peak at the position of the anatase $A_{1\text{g}}$ phonon mode (denoted with \#) was not visible during heating. Second, the final phase of the completed oxidation, depending on whether the laser-induced heating is stopped before or after the rutile phase formation, was rutile or an anatase/rutile mixed-phase (as suggested by Fig. \ref{figOX}b).

\paragraph{Oxidation Kinetics}
The oxidation processes (thermal or photooxidation), induced by gradually increasing the laser power, clearly indicate that HF-etched and LiF–HCl-etched samples stabilize different TiO$_2$ phases -- anatase and rutile, respectively. We remain cautious about attributing the initial oxidation state to anatase, not only due to the unusual blue shift of the $E_{\text{g}}$ band but also because the appearance of the $157$ cm$^{-1}$ peak attests a time- and power-dependent sigmoidal growth under constant laser irradiation. Specifically, this peak grows over $5-15$ minutes at a fixed power and then saturates; once it has stabilized, additional increases in laser power are required to enhance its intensity. Intermediate probe measurements at a lower power of $200~\mu$W ($18$~kW/cm$^2$) performed between power-increment steps, revealed a reduction in the $157$ cm$^{-1}$ peak intensity. However, as the laser power was increased to higher levels, the magnitude of this reduction became less pronounced. The observed decrease—or, in some cases, disappearance—of the $157$ cm$^{-1}$ peak upon lowering the laser power suggests that this newly formed phase is unstable in the early stages of oxidation.

\begin{figure}[htp]
    \centering
    \includegraphics[width=1\linewidth]{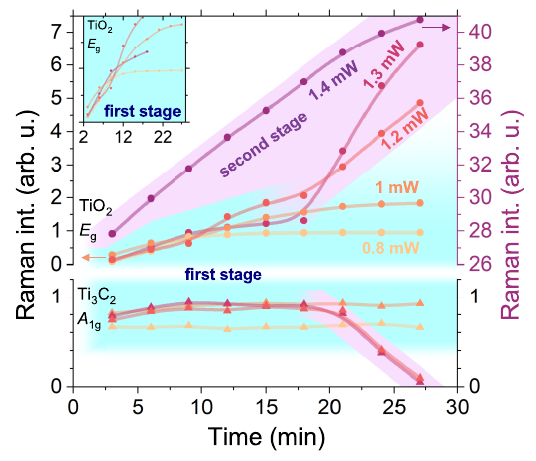}
    \caption{Peak intensities for MX-HF samples as a function of exposure time at a constant laser power. The laser power was varied between $0.8-1.4$ mW ($73-128$~kW/cm$^2$). The first and second stages of the oxidation are highlighted with {\color{cyan}cyan} and {\color{violet}violet} colors, respectively. The inset shows the enlarged region of the first stage with a similar slope and the sigmoidal growth characteristic.}
    \label{figTime}
\end{figure}

A detailed examination of the initial oxidation period is presented in Fig. \ref{figTime} for MX-HF. The response to sustained heating was monitored by following the TiO$_2$ $E_{\text{g}}$ and Ti$_3$C$_2$T$_x$ $A_{1\text{g}}$ Raman peak intensities near $155$ cm$^{-1}$ and $200$ cm$^{-1}$, respectively, under different fixed laser powers. The measurement revealed a two-step oxidation mechanism. The TiO$_2$ $E_{\text{g}}$ peak initially rose, reaching a saturation level indicative of early-stage oxide formation. At a critical power of around $1.3$~mW ($119$~kW/cm$^2$), the growth behavior transitioned to a more rapid, linear increase. The sigmoidal growth observed during the first stage is found to be temperature-independent (see the inset in Fig.\ref{figTime}), displaying the same slope at various temperatures and suggesting diffusion-limited kinetics. At higher laser powers, the anatase-like TiO$_2$ peak growth evolved linearly over time, consistent with zero-order reaction kinetics. The $A_{1\text{g}}$ MXene band (lower panel of Fig. \ref{figTime}) remained unchanged during the diffusion-limited oxide-growth stage; it began to decrease only during the second stage of oxidation. This observation can be explained by the initial formation of a surface oxide that does not disrupt the Ti$_3$C$_2$ structure. In the second stage, oxide formation proceeds between the MXene layers, ultimately degrading the Ti$_3$C$_2$ backbone and reducing the corresponding Raman band intensity.

While the first oxidation stage seems to be temperature-independent with a similar reaction constant, manifested in the temperature-independent slope of the TiO$_2$ $E_{\text{g}}$ peak intensity growth, the temperature increase drives the reaction forward. This cannot be explained with a simple diffusion-controlled reaction. It signals that activation is necessary for the diffusion or for other processes, leading to oxidation. However, the activation temperature cannot be extracted from the \emph{in situ} Raman measurements. Therefore, the so-called cavity perturbation technique \cite{klein_microwave_1993, donovan_microwave_1993} was utilized to obtain more information about the oxidation process by monitoring the microwave conductivity as a function of temperature. This technique measures the shift in frequency regarding the resonance frequency of the empty cavity, $f-f_0$, and the $Q$-factor, $Q=f_0/\Delta f \propto L^{-1}$ of the microwave cavity with and without the presence of the sample. Here, $\Delta f$ is the half-width at half maximum of the cavity resonance response, which is related to microwave loss, $L$.

\begin{figure}[htp]
    \centering
    \includegraphics[width=1\linewidth]{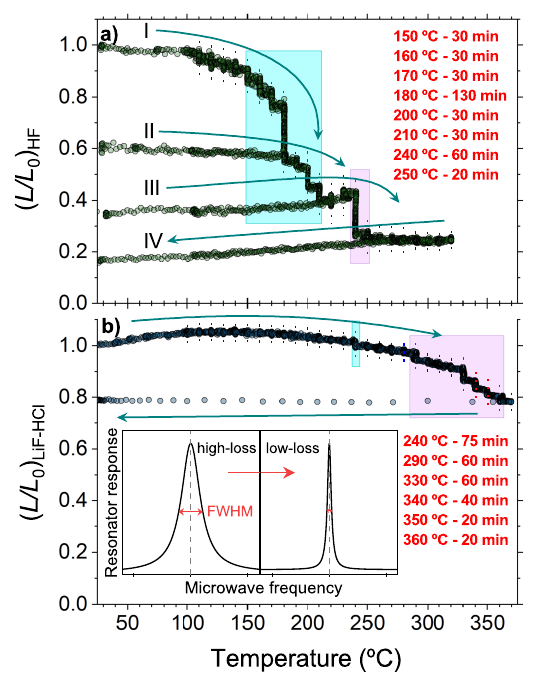}
    \caption{Normalized microwave loss, $L/L_0$, against the temperature of the (a) MX-HF and (b) MX-LiF samples measured by microwave conductivity measurements where $L_0$ is the microwave loss of the pristine material at room temperature. Both samples were heated while open to air. Colored vertical dashed lines show steps, where the temperature was held for a time given in the legend. {\color{red}Red} ({\color{blue}blue}) lines denote where ({\color{blue}no}) significant change is observed. In the case of the HF-etched sample, oxidation begins as low as $150~^\circ$C, and is complete by $260~^\circ$C, whereas the LiF-HCl-etched sample is resistant up to $240~^\circ$C and for a fully oxidized state $360~^\circ$C is required. For the MX-HF sample the experiment was carried out in three heating segments, indicated by Roman numerals, after each the sample was cooled down to room temperature. The final cooldown is labeled as segment IV. {\color{cyan}Cyan} (I) and {\color{violet}violet} (II) shaded areas indicate the two stages of the oxidative transformation. Note the gradual change in the slope indicating that the material is transforming from a mostly metallic state to a doped semiconductor phase (assuming $L\sim \sigma$ justified by literature observations). The bottom inset illustrates the change in the observed microwave resonance induced by changes in the sample.}
    \label{Fig_MWcond}
\end{figure}

As presented in the inset in Fig. \ref{Fig_MWcond}b), the response of the microwave cavity changes in line with a decreased microwave loss. This technique is only sensitive to the conductivity of individual particles as eddy currents are induced on a granular level. In other words, it is not influenced by the resistivity of the grain boundaries and is suitable for detecting small changes in the conductivity of the materials \emph{in situ}.\cite{csosz_giant_2018, markus_tuning_2020} The technique is proven to be a versatile tool to characterize powder materials without the need for physical contacts.\cite{markus_electronic_2018}

The temperature-dependent microwave conductivity measurements performed on MX-HF and MX-LiF are presented in Fig. \ref{Fig_MWcond}. The measurements also showed stepwise transformations in the materials that can be linked to the two-stage oxidation observed in the Raman study. The temperature was continuously increased to $110~^\circ$C followed by $10~^\circ$C increments. At each temperature step, the temperature was held constant for at least $10$ minutes and until no further change in the $Q$-factor was observed before proceeding. In the case of the MX-HF sample, two intermediate cooldowns were also performed to check the conductive behavior between the oxidative stages. The samples were open to air, allowing oxidation to occur. Both materials were stable up to $140~^\circ$C. Above this temperature, the MX-HF material started to oxidize in multiple steps, which can be distinguished into two temperature ranges, similarly to the Raman results. The first stage of oxidation is between $150-210~^\circ$C followed by a second stage between $240-250~^\circ$C. We assumed that the measurements were in the $Q \sim \varrho$ regime throughout the process, which is equivalent to $L \sim \sigma$, where $\sigma$ describes the microwave conductivity of the material. This is supported by previous literature results showing metallic temperature-dependence.\cite{halim_transparent_2014,hart_control_2019,lipatov_metallic_2024} Consequently, the MX-HF sample is transformed from metallic (segments I-II) to a doped semiconductor (segments III-IV) during the multiple step oxidation. No further transformation is observed up to $320~^\circ$C. The color of the material gradually changed from silvery-black to yellowish-white.

In stark contrast, the MX-LiF is resistant up to $240~^\circ$C, where the first stage of oxidation can be observed. The next stage began at $290~^\circ$C and is completed around $360~^\circ$C. Interestingly, apart from the room temperature region ($25-100~^\circ$C), this material did not exhibit a change in the conduction characteristic and remains mostly metallic. A visible change of color was also observed for the MX-LiF sample, however, not as dramatic as in the case of the HF-etched sample. Here, the final product was brownish. These are all probably the results of a difference in surface groups and oxidation pathway underlined by the Raman experiments.

In addition to the temperature-dependent changes in the microwave loss, the cavity perturbation technique also yields the frequency shift with respect to the empty cavity resonance. The frequency shift data shows the same stepwise features, albeit less pronounced, for both materials (not shown) in agreement with the microwave loss results. In conclusion, the microwave conductivity measurements align well with the oxidation behavior observed with Raman spectroscopy. The heat treatments of Ti$_3$C$_2$T$_x$ MXenes cause changes in the material, which occur in multiple steps. Similarly to the laser-induced oxidation, the changes in microwave conductivity can be grouped into two separate regimes. Based on the Raman and microwave conductivity measurements, we propose that the two identified stages correspond to a heat-induced oxidation.

We also note that both materials tend to absorb microwaves, as the at room temperature $Q$-factor is remarkably lowered to about $1{,}150$ (MX-LiF) and to $4{,}500$ (MX-HF), compared to the unloaded cavity with $Q_0~\approx~8{,}000$ (with using the same amount of about $\sim 8$ mg materials for each experiment). Comparing the two, we found that the MX-LiF sample is a gradually better absorber which is in agreement with literature findings.\cite{sengupta_comparative_2020} This underlines the potential use of MXenes as excellent radio-frequency and microwave absorbing materials for a range of applications.

\paragraph{Origin of the Oxide Phase Stabilization}
The oxidation processes (thermal or photooxidation), induced by gradually increasing the laser power, clearly demonstrate that HF-etched and LiF-HCl-etched samples stabilize different TiO$_2$ phases -- anatase and rutile, respectively. Since the main differences among MXenes produced by different synthesis routes lie in their surface terminations, these surface moieties likely play a significant role in driving TiO$_2$ formation. However, it is not an easy task to identify the presence or absence of particular moieties or groups of moieties that promote either anatase or rutile formation during the oxidation of Ti$_3$C$_2$T$_x$. Both XPS and Raman spectroscopy indicate a complex (surface) chemistry for MXenes in general, making peak deconvolution and identification extremely challenging. On the other hand, thanks to the similar behavior of MX-HF 400-Vac, which formed from MX-HF, and MX-LiF, a careful comparison of MX-HF, MX-HF 400-Vac, and MX-LiF can provide the answer to this question. The only difference in the sample preparation between MX-HF and MX-HF 400-Vac is the annealing \emph{in vacuo} at $400~^{\circ}$C, which transforms its oxide affinity from anatase to rutile. MX-LiF, therefore, can be used as a reference for the change. The surface changes after annealing that are similar to that of MX-LiF, can be connected to the stabilization of the different TiO$_2$ phases. Consequently, we investigated the factors responsible for the emergence of these distinct oxide phases.

\begin{figure}[hbp]
    \centering
    \includegraphics[width=1\linewidth]{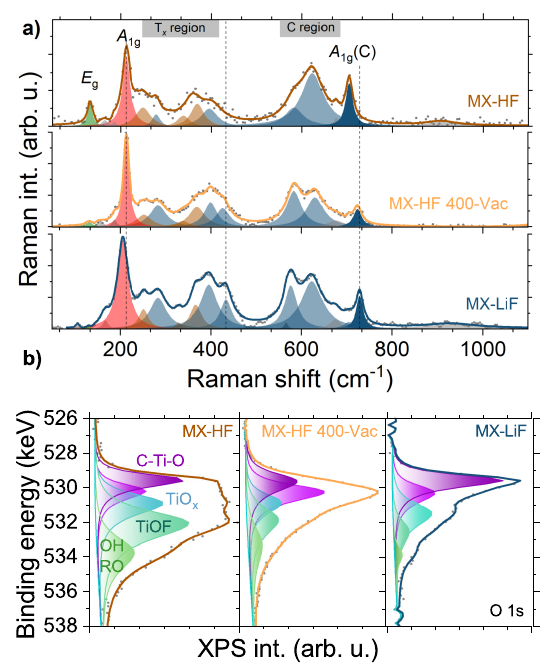}
    \caption{Deconvolution of the as-prepared room temperature Raman spectra of the three different samples ((a) {\color{brown}MX-HF}, {\color{orange}MX-HF 400-Vac}, and {\color{MidnightBlue}MX-LiF}). The Ti$_3$C$_2$T$_x$ $E_{\text{g}}$ and $A_{1\text{g}}$ modes, corresponding to the \ce{Ti\bond{-}C} vibrations, are highlighted in green and red, respectively. Peaks that remained unchanged after vacuum annealing are shown in brown, whereas those that transformed similarly to the MX-HF spectrum are marked in blue. In the deconvoluted O 1s XPS spectra (b) of MX-HF, MX-HF 400-Vac, and MX-LiF, the higher fluorine content in MX-HF is clearly evident.}
    \label{Raman}
\end{figure}

Fig. \ref{Raman}a) presents the Raman spectra of the as-prepared MX-HF and MX-LiF samples and the vacuum-annealed MX-HF 400-Vac. The laser power was fixed to $200~\mu$W ($18$~kW/cm$^2$), and at this excitation power no heating occurs. The recorded broad Raman signals are deconvoluted by using Voigt functions. Numerous studies have attempted to interpret the Raman spectra of MXenes\cite{berger_raman_2023, adomaviciute-grabusove_monitoring_2024, zhang_situ_2018, hu_vibrational_2015, lioi_Electron_2019, sarycheva_raman_2020, sarycheva_tip_2022, limbu_unravelling_2020, plaickner_surface_2024, hu_screening_2018}, and we do not wish to challenge these interpretations here. Nevertheless, in the Raman spectrum of Ti$_3$C$_2$T$_x$, the low-frequency modes at around $130$~cm$^{-1}$ ($E_{\text{g}}$) and $200$~cm$^{-1}$ ($A_{1\text{g}}$) originate from in-plane and out-of-plane vibrations of the outer Ti--C layers, whereas the high-frequency modes at approximately $600$~cm$^{-1}$ ($A_{1\text{g}}$) and $620$~cm$^{-1}$ ($E_{\text{g}}$) are associated with similar vibrational patterns within the central C layers. However, these theoretical zone-center eigenmodes only partially match the experimentally observed Raman spectra. In practice, the aforementioned $E_{\text{g}}$ and $A_{1\text{g}}$ modes, as well as the peak above $700$~cm$^{-1}$, can be linked to zone-center modes of MXenes, whereas the broader features spanning $220-700$~cm$^{-1}$ likely reflect a weighted average of phonons across the entire Brillouin zone. Both the relative abundance and local arrangement of surface species significantly affect the resulting Raman signatures, influencing not only the surface region but also the zone-center modes. 

We labeled in brown the peaks that remain largely unchanged after annealing (peaks in MX-HF was not changed significantly due to the annealing), whereas those that resembled the post-annealing peaks of MX-LiF appear in blue. The Ti$_3$C$_2$-related $E_{\text{g}}$ and $A_{1\text{g}}$ modes are highlighted in green and red, respectively.

The first notable difference that can be observed is a sharp peak at $131$~cm$^{-1}$ in the MX-HF spectrum, which is nearly absent in the other two. The presence of this mode is commonly associated with resonance enhancement. However, because MXenes are metallic, the sample should theoretically always be in a resonant condition. Nevertheless, the decrease in the intensity of the peak linked to the $E_{\text{g}}$ mode is the most pronounced in MX-HF, regardless of the excitation wavelength ($437$, $532$, or $630$~nm). We found that the $A_{1\text{g}}$ mode is at $215$~cm$^{-1}$ for both MX-HF and MX-HF 400-Vac, and is at $205$~cm$^{-1}$ for MX-LiF. Previous studies have reported that the redshift of the $A_{1\text{g}}$ mode in LiF-etched samples correlates with reduced fluorine content \cite{lioi_Electron_2019}. However, as noted above, vacuum annealing also reduces the F concentration but does not significantly harden the $A_{1\text{g}}$ mode, consistent with earlier findings \cite{plaickner_surface_2024}. Though, the surface reconstruction due to the used vacuum-annealing does not affect the $A_{1\text{g}}$ mode.

In addition to the reduced intensity of the $E_{\text{g}}$ mode, the most pronounced alteration is observed in the $A_{1\text{g}}$(C) mode, located at $705$, $725$, and $728$~cm$^{-1}$ in MX-HF, MX-HF 400-Vac, and MX-LiF samples, respectively. Indeed, this peak is known to be highly sensitive to surface terminations, and the observed blue shift can be explained by fluorine depletion and oxygen enrichment at the surface. However, theoretical calculations indicate that substituting \ce{\bond{-}F} with \ce{\bond{-}OH} causes a red shift, while increasing the oxygen termination leads to a blue shift.\cite{hu_vibrational_2015, berger_raman_2023} Despite these nuances, both the experimental and theoretical findings suggest that elevated O termination is responsible for the softening of the $A_{1\text{g}}$(C) mode. It should be noted that, in our deconvolution, all spectra exhibit a peak at $705$~cm$^{-1}$, though this feature is insignificant in the MX-HF 400-Vac and MX-LiF samples.

The Raman region influenced by surface states also showed notable changes. In our measurements, the peak at approximately $620$~cm$^{-1}$ decreases upon annealing, whereas the one at roughly $580$~cm$^{-1}$ is increased, making this region closely resemble that of MX-LiF. This spectral window is generally connected to C vibrations: the former peak is often attributed to OH or F terminations, while the latter is linked to O terminations. In addition, two other OH-related peaks at around $280$ and $430$~cm$^{-1}$ are found to increase after annealing in a vacuum to the intensity similar to that was found in MX-LiF.

Elemental analysis by both energy-dispersive X-ray spectroscopy (EDS, see Supporting Information) and XPS revealed reduced F concentrations after vacuum annealing of HF samples and in LiF-etched samples relative to MX-HF, in agreement with previous studies. Although the deconvoluted Ti~2p XPS spectra showed decreased F-related peaks following vacuum annealing, the difference between MX-HF and MX-LiF is smaller than one might expect based solely on F concentration. In the O~1s region, however, the F-related bonds in MX-HF 400-Vac are significantly diminished to levels comparable to those observed in MX-LiF.

These findings suggest that MX-HF 400-Vac resembles MX-LiF mainly because of its lowered F content, which in MX-HF appears largely bound to oxygen. Hence, the high F concentration (likely in the form of oxyfluorides or \ce{Ti\bond{-}O\bond{-}F} rather than just \ce{C\bond{-}Ti\bond{-}F}) appears to be responsible for stabilizing the anatase phase. Although rutile is the most stable polymorph of TiO$_2$, anatase can exhibit greater stability at the nanoscale due to the interplay between surface and bulk energies, leading to a crossover in thermodynamic stability as particle sizes decrease at around $30$~nm.\cite{levchenko_tio2_2006, lokshin_stabilization_2006} On the other hand, the anatase phase is susceptible to transformation into the rutile phase under certain conditions, which can limit its practical applications. The stabilization of the anatase phase can be achieved through the introduction of fluoride and oxyfluoride compounds \cite{corradini_tuning_2015, yang_anatase_2008, kohlrausch_selective_2021, lokshin_stabilization_2006}. This phenomenon can explain why anatase is often detected at the beginning of the oxidation process and why anatase is the stable oxide phase in MX-HF. When oxide particles are initially small, anatase formation is favored. In an oxyfluorine-rich systems, the presence of \ce{F-}-ions can stabilize the anatase phase even when the particles become larger. Without the stabilization provided by \ce{F-}-ions, the formed oxide stabilizes in the rutile phase.

\section{Summary}

In this study, we systematically investigated the oxidation behavior of Ti$_3$C$_2$T$_x$ MXenes, synthesized via different routes, under laser-induced heating. Our \emph{in situ} Raman and microwave conductivity measurements revealed a complex oxidation pathway governed by the surface chemistry of the MXenes. HF-etched samples exhibit a fluorine-rich surface, leading to the preferential formation of anatase TiO$_2$ at lower temperatures and smaller particle sizes, especially in the presence of fluoride and oxyfluoride species. With increasing temperature, a phase transition to rutile occurs; however, upon cooling, this transition reverses due to the stabilizing effect of \ce{F-} ions. In the absence of oxyfluoride species, which can be achieved by either annealing HF-etched MXenes in \emph{vacuo} or applying LiF-HCl or other synthesis methods resulting in low F content, rutile remains the thermodynamically stable phase. Removing these fluorides notably increases the oxidation resistance and promotes the formation of rutile TiO$_2$.

Our findings underscore the critical influence of surface chemistry and synthesis conditions on the oxidation route and final oxide phase, highlighting the pivotal role played by fluoride and oxyfluoride moieties. Additionally, \emph{in situ} monitoring of the oxidation process revealed a two-step mechanism, characterized by an initially unstable oxide phase with a diffusion-controlled growth. The identification of a two-step oxidation mechanism, with initial surface oxidation followed by bulk oxidation, and the possibility of phase control provides new insights into the engineered synthesis of TiO$_2$@C composites, functionalized MXenes, and other TiO$_2$-based nanostructures from MXenes. We confirmed our findings using microwave conductivity measurements that showed the two-stage oxidation and gave approximate temperature ranges attributed to these stages.

The differences in the oxidation mechanisms between HF- and LiF-HCl-etched MXenes underscore both the similarities and distinctions in their surface chemistry. The presence of oxyfluorides alters the oxidation behavior of the material and can impact various surface modification strategies. Our study showed that through vacuum annealing, a HF-etched sample can be further modified to have a surface resembling that of the LiF-HCl-etched Ti$_3$C$_2$T$_x$, improving its stability against oxidation.% This finding enables researchers with the flexibility to prepare the material in diverse ways.

\section*{Methods}

X-ray diffraction (XRD) diffractograms were obtained using a Bruker D8 Discover high-resolution spectrometer, equipped with a Cu K$\alpha$ line, in the $\theta-2\theta$ geometry.

Scanning electron microscopy (SEM) images were captured in vacuum using a FEI Magellan 400 microscope operated at $10$ kV electrons in the Integrated Imaging Facility (NDIIF) at the University of Notre Dame.

X-ray photoelectron spectroscopy (XPS) measurements were performed at the Materials Characterization Facility (MCF) at the University of Notre Dame, using a PHI 5000 Versa Probe II equipped with monochromatic Al K$\alpha$ X-rays ($15$ kV, $25$ W).

Raman measurements were performed using a WITec alpha300R confocal Raman microscope using a Zeiss LD EC Epiplan-Neofluar $50\times$/$0.55$ long working distance objective and with a $1800$ lines/mm grating.

Microwave conductivity measurements were carried out using the cavity perturbation technique \cite{klein_microwave_1993, donovan_microwave_1993} in the transmission geometry utilizing a standard TE$_{011}$ cylindrical cavity. An HP 83752B microwave sweeper was used as a source of radiation with the microwave power set to $10$ dBm ($10$ mW). The transmitted microwave power is detected using an HP 8472A crystal detector and acquired using a Tektronix TDS-320 oscilloscope. Every point is calculated from a $96-256$ average of rapid scans over the resonance of the microwave resonator. The mass of the samples was around $8$ mg.

\section*{Data availability}

The data needed to evaluate and reproduce the conclusions are present in the paper. Additional data related to this paper are available from the corresponding author upon request.

\section*{References}
\bibliographystyle{wiley-chemistry}
\bibliography{References}

\section*{Acknowledgements}

The authors acknowledge the use of the Notre Dame Integrated Imaging Facility (NDIIF) and Materials Characterization Facility (MCF). F.S. acknowledges the National Research, Development and Innovation Office of Hungary (NKFIH) Grants Nr. K137852, K149457, TKP2021-EGA-02, TKP2021-NVA-02, and 2022-2.1.1-NL-2022-00004.

\section*{Competing interests}

The authors declare no competing interests.

\section{TOC figure}

\begin{figure}[htp]
    \centering
    \includegraphics[width=\linewidth]{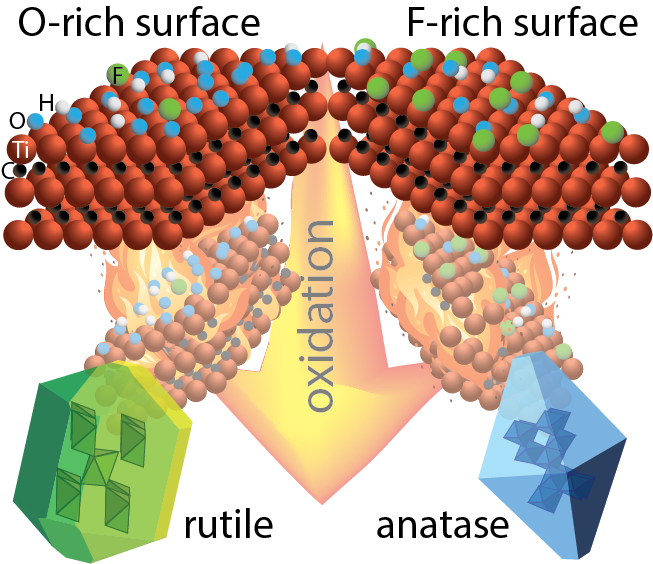}
    \caption{TOC figure.}
    \label{TOC}
\end{figure}

\section{TOC text}
The oxidation of Ti$_3$C$_2$T$_x$ MXenes follows a two-step process, which starts with diffusion-controlled interlayer oxidation and exhibiting surface-dependent characteristics. A fluoride-rich surface facilitates the formation of anatase TiO$_2$, while an oxygen-rich surface favors the development of rutile.

\appendix
\clearpage
\pagebreak
\makeatletter 
\renewcommand{\thefigure}{S\@arabic\c@figure}
\makeatother
\setcounter{figure}{0}

\section{Elemental composition}

\begin{figure}[H]
    \centering
    \includegraphics[width=\linewidth]{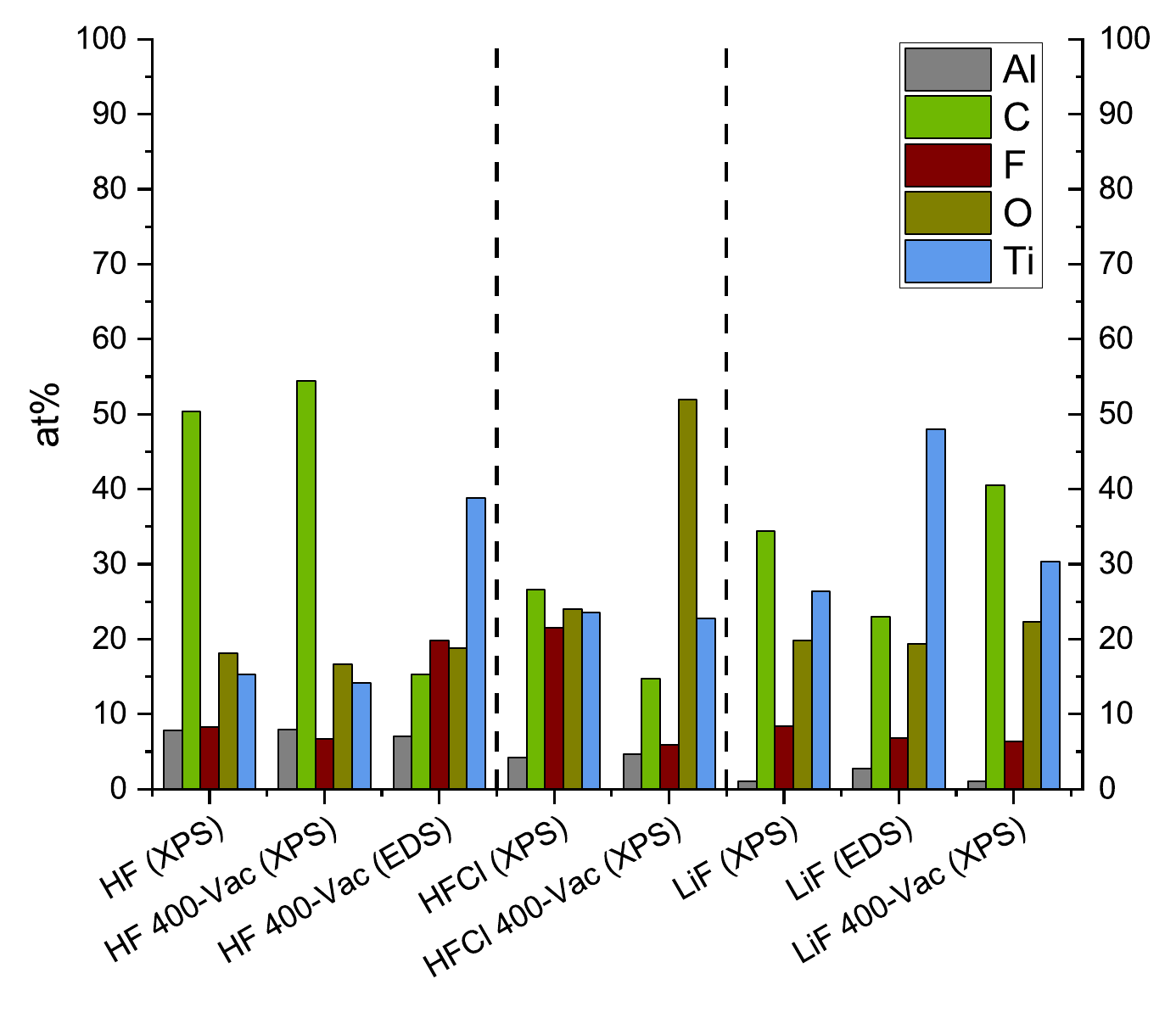}
    \caption{Elemental composition of the investigated materials using XPS (surface sensitive) and EDS (bulk sensitive).}
    \label{FigSI_XPSEDS}
\end{figure}

This Supporting Information provides information about the elemental composition of the prepared materials, as summarized Table \ref{tab:xps_eds_concentration}. and depicted in Fig. \ref{FigSI_XPSEDS}. Two techniques are used to determine the composition: XPS, which is more surface sensitive, and EDS, which is more bulk sensitive.
\begin{table}[H]
    \centering
    \begin{tabular}{l|ccccc|c}
        \hline
        Sample & Al & C & F & O & Ti & Method \\ \hline
        MX-HF & 7.86 & 50.39 & 8.25 & 18.18 & 15.33 & XPS \\
        MX-HF 400-Vac & 7.94 & 54.51 & 6.67 & 16.72 & 14.16 & XPS \\
        MX-HF 400-Vac & 7.08 & 15.36 & 19.89 & 18.87 & 38.8 & EDS \\ \hline
        MX-HFCl & 4.19 & 26.60 & 21.58 & 24.05 & 23.60 & XPS \\
        MX-HFCl 400-Vac & 4.63 & 14.73 & 5.92 & 51.95 & 22.77 & XPS \\ \hline
        MX-LiF& 1.09 & 34.47 & 8.36 & 19.83 & 26.43 & XPS \\
        MX-LiF& 2.73 & 23.00 & 6.81 & 19.40 & 47.99 & EDS \\ 
        MX-LiF 400-Vac & 1.03 & 40.54 & 6.40 & 22.37 & 30.38 & XPS \\ \hline
    \end{tabular}
    \caption{Elemental composition (in at$\%$) obtained from XPS and EDS measurements. XPS results are listed first for each sample, followed by EDS results (when available).}
    \label{tab:xps_eds_concentration}
\end{table}
\vspace{-5mm}

Tables \ref{tab:ti2p_peak_contribution}. and \ref{tab:o1s_peak_contribution}. summarize contributions from different species of titanium and oxygen using XPS after deconvolution.

\begin{table}[H]
    \centering
    \begin{tabular}{l|cccc}
        \hline
        Ti 2p peak & MX-HF& MX-HF 400-Vac & MX-LiF & MX-LiF 400-Vac \\ \hline
        Ti I   & 13.0  & 12.0 & 11.0 & 13.0  \\
        Ti II  & 18.0  & 13.0 & 21.0 & 25.5 \\
        Ti III & 9.7   & 1.2  & 9.4 & 13.7  \\
        Ti IV  & 3.0   & 5.0  & 12.7 & 2.0 \\ \hline
        Ti$^{3+}$ & 3.0 & 9.7 & 1.5  & 3.8  \\
        Ti$^{2+}$ & 8.0 & 7.5 & 4.0 & 3.7   \\ \hline
        TiO$_2$ & 8.5   & 5.0 & 7.0 & 3.7   \\
        TiO$_x$F   & 7.5   & 4.1 & 3.89 & 1.17 \\ \hline
    \end{tabular}
    \caption{Peak contribution in different samples ($\%$ concentration) for the Ti 2p XPS peaks.}
    \label{tab:ti2p_peak_contribution}
\end{table}

\begin{table}[H]
    \centering
    \begin{tabular}{l|cccc}
        \hline
        O 1s Peak & MX-HF & MX-HF 400-Vac & MX-LiF & MX-LiF 400-Vac \\ \hline
        \ce{C\bond{-}Ti\bond{-}O} I & 16.17 & 21.73 & 39.68 & 21.34 \\
        \ce{C\bond{-}Ti\bond{-}O} II & 10.92 & 30.35 & 20.90 & 42.16 \\ \hline
        \textbf{TiOx} & \textbf{19.63} & \textbf{15.68} & \textbf{25.31} & \textbf{21.33} \\
        \textbf{TiOF} & \textbf{36.27} & \textbf{19.65} & \textbf{10.62}  & \textbf{9.97} \\ \hline
        OH / \ce{R\bond{-}O} & 17.01 & 12.59 & 3.49 & 4.60  \\ \hline
    \end{tabular}
    \caption{Peak contribution in different samples ($\%$ concentration) for the O 1s XPS peaks. Rows for \textbf{TiOx} and \textbf{TiOF} are highlighted in bold.}
    \label{tab:o1s_peak_contribution}
\end{table}

\end{document}